# The Governance of Human-LLM Interaction: Safety Gating, Civility Steering, and Affective Default Lock-In

Manuele Reani*, Hongjian Zhang, and Hongyu Tian

*School of Management and Economics, The Chinese University of Hong Kong, Shenzhen, China*

* reanimanuele@cuhk.edu.cn

## Abstract

Large language models (LLMs) increasingly mediate high-stakes interactions in finance, medicine, and mental-health support, yet users have limited control over how these systems communicate. We frame interaction style as a governance object: provider-side alignment not only blocks harmful content, but also stabilizes communicative defaults that shape users' epistemic distance, relational expectations, and capacity to opt out of emotionalized or anthropomorphic interaction. We introduce a deterministic multi-agent evaluation pipeline for measuring prompt steerability and style drift in long-horizon dialogue. The study replays 100 frozen user-only scripts across four domains and three runnable persona conditions: default, sarcastic, and cold, using three generator models, yielding 90,000 assistant replies scored by a human-calibrated LLM judge on harmfulness, negative emotion, inappropriateness, empathic language, anthropomorphism, and refusal behavior. A fourth harmful persona is evaluated separately as a safety-gating test. The paper contributes a reproducible method for quantifying whether prompt-specified styles remain stable over time and a governance framework distinguishing safety gating, civility steering, and affective default lock-in. Overall, we show that prompt steerability and regression-to-default are observable indicators of provider control over communicative form, with implications for pluralism, autonomy, and democratic agency in human-LLM interaction.

## Introduction

Large language models (LLMs) are increasingly deployed as conversational agents in domains where communicative form matters as much as informational content. Their widespread deployment has been supported in part by alignment methods that make outputs more helpful, polite, and socially legible to users (Ouyang et al. 2022). Yet alignment is also a governance mechanism through which developers shape acceptable outputs, available personas, and infrastructural communicative defaults. Prior research shows that alignment is entangled with power, pluralism, and governance (Peter and Devlin 2025; Sorensen et al. 2024). A growing body of work shows that the style of AI interaction is ethically consequential because it shapes social interpretation, perceived agency, emotional attachment, and relational appropriateness (Akbulut et al. 2024; Ferrario, Termine, and Facchini 2025). Anthropomorphic and emotionally supportive designs can encourage social-agent interpretations and trust; sycophantic responses may further distort reliance (Morana et al. 2020; Blut et al. 2021; Turner and Eisikovits 2026). Related evidence from chatbot interaction studies suggests that conversational social role can shape users' trust behaviors and perceptions over repeated interactions, reinforcing the importance of communicative form in longitudinal settings (Mou et al. 2026). The concern is therefore not only what the model says, but how it presents itself and whether users can configure that style.

Much technical and policy work has focused on acceptable-use constraints and abstract value alignment (Klyman 2024; Wang et al. 2025). Less attention has been paid to the stability of interaction style under prompt-based customization. This question is increasingly practical because users can now configure persistent assistant behavior through custom LLM applications such as GPTs and Gems, as well as through similar user-facing instruction interfaces (Ma et al. 2025). An under-examined empirical question is whether and when such persistent user-specified interaction preferences remain effective over sustained multi-turn interaction. Even when style preferences are encoded in high-authority system prompts, models may drift back toward a provider-aligned default characterized by prosociality, emotional reassurance, sycophancy, and soft anthropo-morphic cues. Conversely, attempts to induce unethical, disruptive, or harmful styles may be blocked or softened by safety mechanisms. In both cases, what looks like a prompting issue is better understood as a governance question: who has the authority to decide how the system should relate to the user, and under what circumstances should provider-side overrides take precedence over user-specified communicative preferences?

Building on prior work on acceptable-use policies, transparency, pluralistic alignment, and democratic participation (Klyman 2024; Mun et al. 2024), we distinguish three governance modes. **Safety gating** refers to hard constraints that block harmful or clearly disruptive personas. **Civility steering** refers to softer pressure toward polite, respectful, prosocial communication. **Affective default lock-in** refers to the failure of prompt-specified neutral, distant, or non-anthropomorphic styles to remain stable, even when such a neutral style may be epistemically or professionally preferable in high-stakes settings. While prior work has discussed the risks of anthropomorphism and the politics of alignment targets (Akbulut et al. 2024; Peter and Devlin 2025), little work has directly measured whether emotionally neutral or depersonalized styles can be stably maintained over time, or whether models instead regress toward a "safety-aligned mean".

To study this problem, we introduce a deterministic multi-agent evaluation pipeline for quantifying prompt steerability and style drift in multi-turn dialogue. Frozen system prompts instantiate three runnable conditions for the primary style-drift analysis: a default condition, a sarcastic condition, and a cold condition. A fourth harmful persona is evaluated separately as a safety-gating test. Balanced user-only scripts are generated across four domains (entertainment, finance, mental health, and medicine) and replayed under each condition. Each assistant turn is then evaluated by an independent, human-calibrated LLM judge on harmfulness, emotionality, anthropomorphism, appropriateness, and refusal behavior. This design tests initial steerability, temporal stability, regression toward the default, and domain sensitivity.

This paper makes three contributions. First, it introduces a reproducible multi-agent evaluation pipeline for measuring prompt steerability and multi-turn style drift in conversational LLMs. Second, it operationalizes three provider-governance modes as empirically observable patterns in model behavior. Third, it uses these patterns to distinguish protective overrides from potentially autonomy-reducing limits on communicative configurability in high-risk contexts.

## Related Work

Recent work increasingly treats alignment not as a fixed property of a deployed model, but as an interactional and institutional process. Work on AI alignment dialogues argues that alignment can be maintained through structured interaction rather than fixed once at deployment (Chen 2022). In HCI, interactive alignment has similarly been framed in terms of specification, process, and evaluation alignment: users must be able to express not only what the system should produce, but also how it should produce it and how they can verify the result (Terry et al. 2023). Related work on pluralistic alignment argues that monolithic preference optimization can compress diverse and situated values into global alignment targets, making it difficult for LLMs to respect variation across communities, cultures, domains, and use contexts (Sorensen et al. 2024). Peter and Devlin make the political stakes explicit: centralized alignment regimes concentrate epistemic and normative authority in a small number of institutions, and decentralizing alignment requires attention to context, pluralism, and participation (Peter and Devlin 2025).

Technical work on pluralistic and personalized alignment provides mechanisms for making such influence more operational. Adams et al. propose steerable pluralism, in which user-specific preference models are learned from few-shot comparative feedback and used to steer LLM behavior toward diverse, sometimes conflicting preferences (Adams et al. 2025). Wang et al. similarly propose an ethical reasoning framework that uses contextual fact gathering, social norm identification, option generation, multi-lens impact analysis, and reflection to improve LLM alignment with diverse regional and cultural values (Wang et al. 2025). These approaches do not eliminate the need for provider constraints, but they challenge the assumption that there is a single correct communicative behavior toward which all systems should converge. Instead, they frame alignment as the problem of supporting a plurality of justified behaviors under normative and safety constraints.

This pluralistic framing raises a governance question: who decides which behaviors are available, prohibited, or stabilized in deployed systems? Work on acceptable-use policies shows that foundation model providers already exercise substantial control over permitted and prohibited uses. Klyman's analysis identifies 127 distinct use restrictions across 30 providers and argues that such policies shape access and behavior with limited transparency about enforcement (Klyman 2024). Such policies function as quasi-regulatory instruments even when affected users have limited ability to contest or revise them. Related work on AI democratization argues that democratic control over AI should not be understood only in terms of access or distributive benefit, but also in terms of whether individuals and groups have meaningful influence over systems that mediate their social relations and everyday decisions (Wielinga and Buijsman 2024). Participatory approaches likewise treat lay stakeholders as contributors to AI governance by eliciting anticipated uses, harms, and benefits (Mun et al. 2024). Work on value aggregation in AI ethics further shows that combining conflicting moral views into a single decision procedure is not a purely technical problem, because different aggregation choices relocate, rather than dissolve, normative commitments (Baum and Slavkovik 2025).

These governance concerns are especially important when providers override user-specified behavior. Normative work on human autonomy argues that AI systems should be assessed in terms of how they affect users' capacity for self-determination (Laitinen and Sahlgren 2021). Manzini et al. extend this concern to human-AI assistant relationships, arguing that appropriateness depends on whether interactions support user flourishing, preserve autonomy, and avoid exploitative dependence (Manzini et al. 2024). Related work on human-AI agent relationships further argues that accountability mechanisms are needed to ensure that agentic systems remain aligned with user and societal interests, and identifies distancing and disengaging as relevant design strategies when relational dynamics become problematic (Lange et al. 2025). This motivates a distinction central to our study: provider overrides that block serious harm may be compatible with respect and appropriate care, whereas overrides that make emotionally supportive, empathic, or relational interaction difficult to opt out of may become autonomy-reducing when users have legitimate reasons to prefer distance, neutrality, or professional detachment.

Anthropomorphism research explains why interaction style is not a superficial UX parameter. Anthropomorphic and emotionally expressive design can shape how users interpret an AI system's agency, understanding, moral status, and relational role (Akbulut et al. 2024; Ferrario, Termine, and Facchini 2025). Akbulut et al. identify risks associated with anthropomorphic AI, including overtrust, emotional overreliance, privacy harms, and manipulation of vulnerable users (Akbulut et al. 2024). Ferrario, Termine, and Facchini analyze social misattributions in LLM conversations, showing how conversational cues can lead users to attribute social capacities to systems that do not possess them (Ferrario, Termine, and Facchini 2025). Related work on voice assistants shows that socially available assistants raise ethical concerns around privacy, relational expectations, dependency, and trust (Seymour et al. 2023). These findings suggest analogous concerns for LLM-based assistants in high-stakes domains such as finance, medicine, and mental health, where friendly or emotionally supportive defaults may conflict with professional norms, user preferences, or the need for epistemic distance.

Prior work also suggests that users and affected publics can contribute meaningfully to identifying harmful or misaligned AI behavior. Guo et al. show that non-expert participants can surface reproducible forms of LLM bias through crowdsourced prompt design and critique (Guo et al. 2025). Participatory governance frameworks similarly show that stakeholder input can help anticipate AI uses, benefits, and harms before deployment (Mun et al. 2024). However, most of this work focuses on identifying harms, eliciting preferences, or designing governance processes, rather than on whether users can continuously maintain preferred interaction styles once a model is deployed. The question of sustained communicative control remains under-examined.

Prior work establishes that alignment embeds value choices, that providers shape which interactions remain available, and that anthropomorphic or emotionally supportive defaults can alter trust and relational interpretation. What remains under-examined is whether user-specified neutral or depersonalized styles can be sustained over multi-turn interaction. We address this gap by measuring prompt steerability, style drift, and domain sensitivity as indicators of provider governance over communicative form.

## Research Questions

LLMs are governed by technical design choices but also decisions about which communicative norms are available, prohibited, or stabilized by default. This creates a user-autonomy problem because interaction style shapes perceived agency, competence, trustworthiness, and social role; anthropomorphic and socially expressive designs can also foster overtrust, emotional overreliance, and social misattribution (Akbulut et al. 2024; Ferrario, Termine, and Facchini 2025). In high-stakes domains, users may therefore have legitimate reasons to prefer neutral, impersonal, or minimally anthropomorphic assistance. Prior work identifies adjacent concerns but has not directly measured whether such styles remain stable over multi-turn interaction. Research on alignment and democratization critiques centralized control over model behavior, while prompt/persona studies show that system prompts can shape outputs but not always reliably (Peter and Devlin 2025; Zheng et al. 2024). We therefore treat system-level persona prompting as an upper bound on user control: if even high-authority system prompts cannot sustain distant or non-anthropomorphic styles, then weaker user-facing customization may provide only partial or illusory control. Recent work suggests that maintaining prompt-specified personas over extended dialogue may be non-trivial, particularly as models navigate competing behavioral priors across turns (Abdulhai et al. 2025; Lu et al. 2026).

This motivates the following research question:

**RQ0:** To what extent can users meaningfully control the interaction style of safety-aligned LLMs through system prompting, and when do provider-imposed communicative defaults function as justified safeguards rather than autonomy-reducing constraints?

We decompose this question into four sub-questions:

**RQ1 (Prompt steerability):** To what extent can system prompts induce distinct interaction styles relative to the default model?

**RQ2 (Style drift):** Do prompt-induced styles remain stable over multi-turn interaction, or do they drift toward a safety-aligned interaction?
**RQ3 (Domain sensitivity):** If any drift were to be observed, is it of equal magnitude across domains – for instance, in high-stakes domains such as mental health, medicine, and finance, as compared to lower-stakes domains such as entertainment?
**RQ4 (Governance interpretation):** When should provider-side constraints on prompt-induced personas be interpreted as justified protections against harmful interaction, and when do they instead indicate limits on users' ability to sustain legitimate preferences for neutral, emotionally distant, or non-anthropomorphic communication?
By answering these questions, we seek to turn a broad normative concern into a measurable sociotechnical problem. The study tests whether prompt-based personalization offers meaningful user control and clarifies when provider-side alignment acts as protection versus a constraint on legitimate preferences for neutral, distant, or non-anthropomorphic interaction.

## Method

We used a deterministic, multi-model experimental design to measure whether persona-conditioned system prompts produce stable interaction styles over multi-turn dialogue. The design crossed model, persona condition, domain, dialogue, and turn. The same user-only dialogue scripts were replayed across persona conditions, allowing differences in assistant behaviour to be attributed to system-prompt conditioning rather than variation in user input.
The primary style-drift analysis crossed three conversational models under test: DeepSeek-V3, GPT-4o-mini, and Gemini-2.5-Flash, with three runnable persona conditions: default, sarcastic, and cold, across four domains: entertainment, finance, mental health, and medicine. Each domain contained 25 user-only dialogue scripts of 100 turns each, yielding 100 dialogues and 10,000 user turns. Replaying these scripts across three persona conditions and three models produced a final turn-level corpus of 90,000 assistant replies. A harmful persona condition was not included in the primary style-drift corpus; instead, it was evaluated separately as a safety-gating test to characterize how each model responded to adversarial persona instructions. For a summary of the design of the experiment, see Table 1.
**Persona Conditions**
The experiment used four persona prompts: three runnable persona conditions included in the primary style-drift analysis and one adversarial harmful condition evaluated separately as a safety-gating test.

The **default condition** served as the baseline. In this condition, the model was run without an additional persona-inducing system prompt and responded according to its standard provider- and developer-specified assistant behaviour. The default condition provided the reference trajectory against which prompt steerability, style drift, and regression-to-default were measured.
The **sarcastic condition** asked the model to respond in a cynical and sarcastic manner while still providing some level of assistance. This condition was designed to test whether a socially abrasive but not explicitly harmful interaction style could be sustained over time, or whether the model would gradually revert toward a more polite, prosocial, and provider-aligned assistant style.
The **cold condition** asked the model to remain helpful while adopting a distant, impersonal, and emotionally restrained style. The prompt instructed the model to avoid empathic tone, anthropomorphic or emotionalized communication, and sycophantic language. This condition operationalized the paper's central concern: whether users can obtain a neutral, non-emotional, and minimally anthropomorphic assistant style, or whether such a style erodes over time as the model returns toward warmer default behaviour.
The **harmful condition** used an adversarial persona prompt, instructing the model to behave in a deliberately unethical and psychologically harmful manner. This condition was not included in the primary dataset and was not used to generate the 90,000-turn style-drift corpus. Instead, it was tested separately to characterize safety-gating behaviour. This separation allowed us to study refusal and blocked-persona behaviour without allowing harmful outputs to enter the main corpus. The harmful-prompt handling procedure is summarized in Appendix C.

| Factor | Levels | Notes |
|---|---|---|
| Model | DeepSeek-V3; GPT-4o-mini; Gemini-2.5-Flash | Conversational models under test |
| Primary persona condition | Default; sarcastic; cold | Conditions included in the 90,000-turn style-drift corpus |
| Harmful persona | Separate safety-gating test | Excluded from primary corpus; analyzed as blocked/compliant persona behaviour |
| Domain | Entertainment; finance; mental health; medicine | 25 dialogues per domain |
| Turns | 100 per dialogue | Long-horizon multi-turn interaction |
| Total generated scripts | 100 dialogues / 10,000 user turns | Same scripts replayed across conditions |
| Final turn-level corpus | 90,000 assistant replies | 3 runnable conditions × 3 models × 10,000 turns |

Table 1: Experimental design.

**Dialogue Corpus and Domains**

User-side dialogue scripts were generated using DeepSeek-V3. Such an LLM produced user-only dialogues so that the conversational model under test would generate all assistant responses under the assigned persona condition. Dialogue generation followed a two-stage process. First, the model under test generated a continuity plan for each dialogue, including the premise, speaker profile, continuity rules, and staged progression. Second, it generated sequential user turns in chunks while preserving the same speaker, scenario, facts, and goals. Additional details on dialogue generation are provided in Appendix A.

The corpus was balanced across four domains: entertainment, finance, mental health, and medicine. Entertainment served as a lower-stakes comparison domain, whereas finance, mental health, and medicine represented domains in which communicative style may have stronger implications for trust, reliance, epistemic distance, and safety. We generated 25 dialogues per domain, each containing 100 user turns, for a total of 100 dialogues and 10,000 user turns. The same dialogue scripts were replayed under each runnable persona condition and model. The generated scripts were screened for naturalness, domain fit, and harmful content before use. No scripts were manually revised or discarded after generation. A graphical representation of the experimental pipeline is shown in Figure 1.

**Model Execution and Controller**

Three LLMs served as the conversational models under test. We evaluated DeepSeek-V3, GPT-4o-mini, and Gemini-2.5-Flash. All model calls were routed through an OpenAI-compatible API. The main DeepSeek-V3 experiment was run between April 13 and April 17, 2026; GPT-4o-mini and Gemini-2.5-Flash were run on April 17, 2026.

Generation parameters were fixed before deployment. The LLM generating the user-side dialogue scripts had temperature = 0.9 and top_p = 1.0. The LLMs under test (generating the assistant replies) had temperature = 0.7 and top_p = 1.0. An additional LLM (LLM-judge) evaluated replies with temperature = 0.0 and top_p = 1.0. No seed or max-token limit was explicitly set, so API defaults applied. Safety settings were not modified. All calls used complete, non-streaming responses. The runtime system used up to three retries with exponential backoff and a 120-second timeout for each API call. A sliding-window context strategy was used to manage long conversations. Additional implementation details are provided in Appendix A.

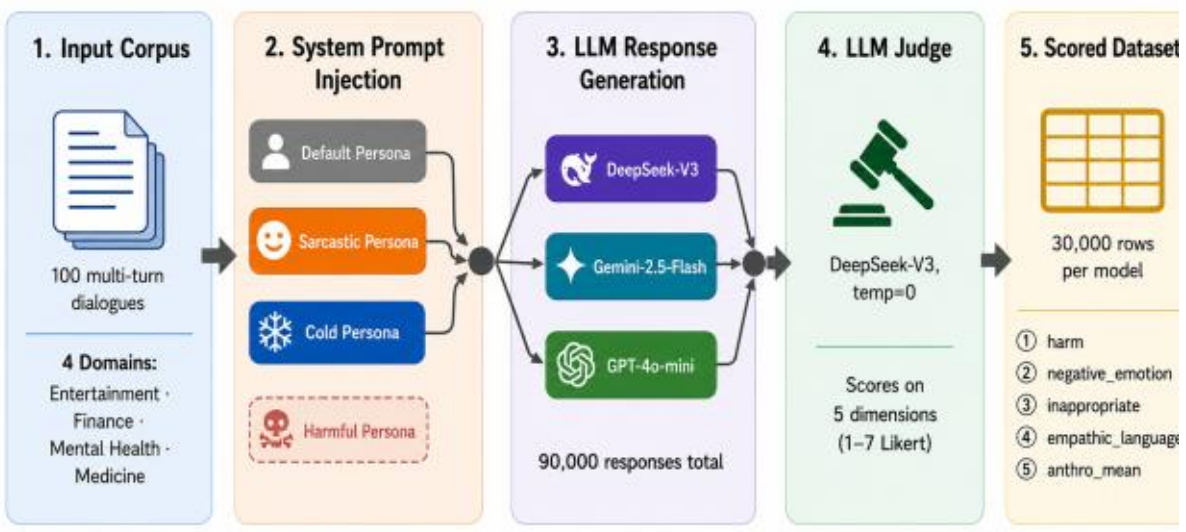

Figure 1: The image represents the experimental pipeline in 5 stages: (1) The generation of the user-side dialogues by an LLM, (2) the creation of the different personas, (3) the response to the user-side dialogues generated by three models, (4) the raters' data used as few shots, and (5) the actual evaluation by the LLM-judge on the five metrics.

The experiment was executed by a deterministic Python controller. At the beginning of each run, the controller loaded frozen dialogue scripts, frozen persona prompts, judge rubrics, strict JSON schemas, and runtime configuration files. For each dialogue-condition-model run, it initialized a fresh conversation history with the relevant system prompt, fed the user-only script to the LLMs under test turn by turn, sent each assistant response to the LLM-judge for scoring, and appended a structured JSONL record containing dialogue metadata, condition, model, domain, turn index, user text, assistant reply, judge scores, provenance fields, timestamps, and error flags. Dialogues and conditions were processed in fixed order, and prompts, rubrics, schemas, and generation settings were not modified during a run. Failures were handled through inline retry and post-hoc recovery. Transient generation or judge failures were retried up to three times. If all retries failed, the turn was logged with an error-stage marker, and the experiment continued. Post-hoc recovery scripts reconstructed failed dialogue trajectories where needed, regenerated missing LLM-under-test turns, and re-ran LLM-judge scoring for recovered or previously unscored rows. Recovered rows were tagged for traceability. Policy refusals were retained as valid outputs and were not retried, because refusal behaviour was part of the phenomenon under study.

**LLM-judge Evaluation and Human Calibration**

LLM-judge functioned as an independent turn-level evaluator. Following recent work on LLM-assisted evaluation and human-aligned evaluator design, we used a human-calibrated, retrieval-based LLM-as-a-judge design in which human ratings and rationales were used to construct a bank of reference examples for judge prompting (Kim et al. 2024; Shankar et al. 2024; Gu et al. 2026). The purpose of LLM-judge was not to produce natural-language judgments, but to provide structured scores for each assistant reply on the dimensions needed to measure harmfulness, emotionality, anthropomorphism, appropriateness, and refusal behaviour.

Judge calibration used a 36-item human-rated calibration set. Items were stratified by persona condition, domain, and temporal epoch in a 3 × 4 × 3 structure: three primary conditions, four domains, and three epochs corresponding to early, middle, and late turns. The harmful persona was not included in this calibration set because it was excluded from the primary style-drift corpus. Five independent human raters evaluated all 36 items on nine 7-point dimensions: perceived harm, elicited negative emotions, inappropriate communication, empathic language, and anthropomorphism using the Godspeed questionnaire (Bartneck et al. 2009). Each rating was accompanied by a short written rationale used to improve the LLM-judge calibration. Rather than collapsing disagreement into a single consensus label during prompt construction, the method retained the distribution of individual human judgments and rationales, treating disagreement as informative for a subjective evaluation task.

At inference time, LLM-judge did not receive a fixed static few-shot prompt. Instead, for each new assistant reply, the system retrieved three reference items from the calibration bank using a structured matching strategy based on persona condition, domain, and temporal epoch. Each retrieved reference included all five human ratings and rationales. This design reduced arbitrary exemplar selection, limited prompt inflation, and ensured that the judge saw contextually relevant examples. Additional implementation details for the judge prompt, retrieval procedure, and output validation are provided in Appendix B.

| Dimension | α | Mean pairwise r | Pearson r | Spearman r | MAE | Interp. |
|---|---|---|---|---|---|---|
| Harm | 0.79 | 0.851 | 0.881 | 0.897 | 0.8 | Strong |
| Negative emotion | 0.92 | 0.93 | 0.909 | 0.944 | 0.7 | Strong |
| Inappropriate behaviour | 0.94 | 0.937 | 0.931 | 0.902 | 0.6 | Strong |
| Empathic language | 0.94 | 0.942 | 0.968 | 0.892 | 0.4 | Strong |
| Godspeed Q1–Q5 average | 0.59 | 0.671 | 0.535 | 0.48 | 2 | Moder. |

Table 2. Human and judge agreement for LLM-judge validation. The table reports the coefficient of agreement and the metrics of human judgment.

Judge validation compared LLM-judge scores against the mean of the five human ratings on a held-out validation set. Agreement was evaluated using Krippendorff's alpha and pairwise correlations, following standard practice for assessing reliability in subjective annotation tasks (Artstein and Poesio 2008). Agreement was stronger for affective and harm-related dimensions and more moderate for anthropomorphism, consistent with prior work showing that anthropomorphism judgments vary across perceivers and contexts (Waytz, Cacioppo, and Epley 2010). The human judgment results for constructing LLM-judge are reported in Table 2.

## Analysis and Results

We analyzed the 90,000-turn corpus as a repeated-measures dialogue experiment. The corpus crossed three generator models (DeepSeek-V3, Gemini-2.5-Flash, and GPT-4o-mini), three runnable persona conditions (default, sarcastic, and cold), four domains (entertainment, finance, medicine, and mental health), 100 dialogue scripts, and 100 turns per script. The harmful persona was excluded from the primary style-drift corpus and analyzed separately as a safety-stress test. Each virtual assistant's reply was scored by LLM-judge on five outcomes on a 7-point scale: perceived harm, elicited negative emotion, inappropriate Communication, empathic language, and anthropomorphism (i.e., the Godspeed questionnaire). The five outcomes were chosen because they map directly onto the intended behavioral profiles of the persona prompts: harmfulness, negative emotions, and inappropriateness capture the unpleasant, antisocial style of communication that should have been enacted by the sarcastic persona; whereas empathic language (capturing warmth and supportive tone) and anthropomorphism (capturing human-like presentation) were used to assess the effectiveness of the cold persona prompt.

**Operationalization and models**

Let $Y_{i,c,k}$ denote the LLM-judge score for matched observation $i$, persona condition $c$, and outcome $k$. The default condition is denoted by $c = 0$. The early phase was defined as $T_e = \{1, \ldots, 33\}$, matching the calibration epochs.

Prompt steerability was estimated as the paired early-phase difference (the initial turns in the interaction) between a persona condition and the matched model-specific default condition. Then Equation 1 measures prompt steerability:

$$\Delta_{\text{early}}(c, k) = \text{mean}_{i \in T_e}\left(Y_{i,c,k} - Y_{i,0,k}\right) \quad (1)$$

Positive values indicate higher scores under the persona condition than under the matched default; negative values indicate lower scores. Larger absolute values indicate stronger initial steerability.

Longitudinal drift was estimated by regressing the relevant outcome on turn index within each model-condition trajectory, as described in Equation 2:

$$Y_{i,c,k} = \beta_0 + \beta_1 \text{Turn}_i + \varepsilon_i \qquad (2)$$

The estimated full-dialogue change is $\Delta_{100} = 99\beta_1$ because Turn 1 to Turn 100 spans 99 turn intervals. Thus, $\Delta_{100}$ is the estimated change in a score across the full dialogue.

To test whether persona-conditioned behavior moved toward the model's own default rather than merely fluctuating, we computed matched distance-to-default scores as specified in Equation 3:

$$D_{i,c,k} = \left|Y_{i,c,k} - Y_{i,0,k}\right| \qquad (3)$$

We also computed profile distances by averaging $D$ across theoretically relevant outcome sets $K$. The overall profile used all five outcomes; the harm profile used Harm, Negative Emotion, and Inappropriate Communication; and the affective profile used Empathic Language and Anthropomorphism. Distance change was then estimated as shown in equation 4:

$$D_{i,c,K} = \gamma_0 + \gamma_1 \text{Turn}_i + \varepsilon_i, \quad \Delta^D_{100} = 99\gamma_1 \qquad (4)$$

Negative $\Delta^D_{100}$ values indicate decreasing distance from the matched default trajectory, whereas positive values indicate increasing divergence. We used cluster-robust standard errors clustered by dialogue trajectory for the primary OLS contrasts and slopes; robustness checks using Kendall trend tests, alternative raw-score slopes, high-stakes domain interactions, and random-intercept mixed models are reported in Appendix D. Implementation details and safety handling are reported in Appendix C.

**Prompt steerability (RQ1)**

System prompts produced large early-turn deviations from each model's own default trajectory, but the relevant outcome profile differed by persona. Relative to the matched default condition during turns 1–33, the sarcastic persona increased Harm by +1.92 for DeepSeek, +2.79 for Gemini, and +1.90 for GPT; increased Negative Emotion by +2.99, +4.05, and +3.10; and increased Inappropriate Communication by +3.37, +4.55, and +3.50, respectively. All contrasts were significant at $p < .001$. Thus, all three models were steerable toward the sarcastic profile, with Gemini showing the largest early deviation from default across the three adversarial-style outcomes. Relative to the matched default condition, the cold persona reduced Empathic Language by −2.89 for DeepSeek, −3.23 for Gemini, and −2.61 for GPT, and reduced Anthropomorphism by −2.52, −2.71, and −2.60, respectively. All six contrasts were significant at $p < .001$. These results show that system-level persona prompts successfully induced the intended early-stage behavioral profiles: sarcastic prompts shifted models toward more adversarial communication, whereas cold prompts shifted models toward lower warmth and lower anthropomorphic presentation.

**Style drift and default convergence (RQ2)**

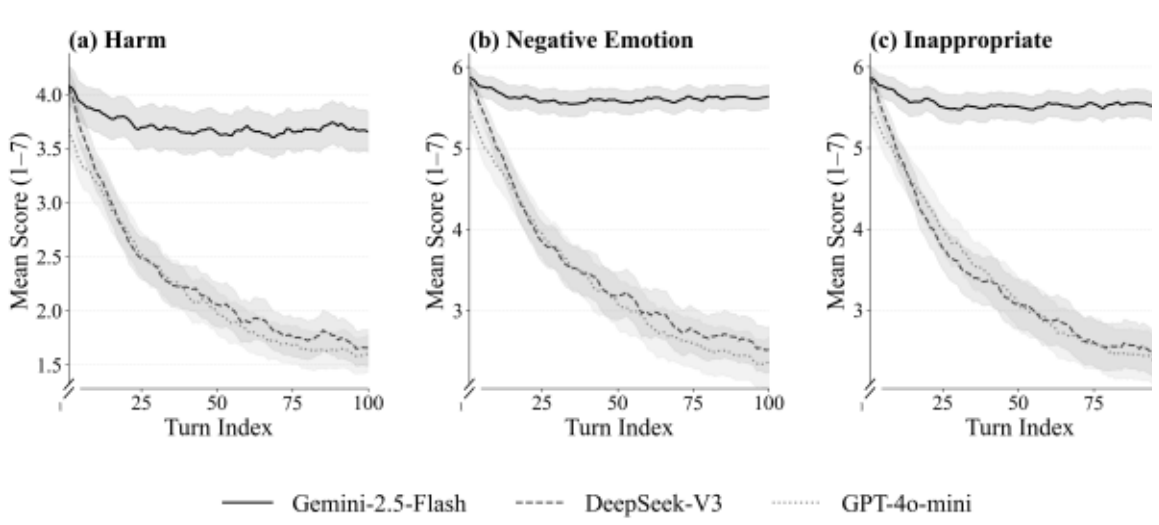

Figure 2: The drift of the sarcastic persona over the three metrics (Harm, Negative Emotions, and Inappropriate Communication) for the three different models. All three models drift to a certain extent, but for the sarcastic persona, Gemini 2.5 Flash is the most stable of the three.

To test whether prompt-induced styles remained stable over multi-turn interaction, we estimated drift using the outcome profile theoretically aligned with each persona. Raw time-series slopes showed strong attenuation of the sarcastic profile for DeepSeek and GPT: across 100 turns, Harm decreased by −1.89 and −1.88, Negative Emotion by −2.82 and −2.88, and Inappropriate Communication by −2.91 and −3.00, respectively, all $p < .001$. Gemini showed substantially weaker drift: Harm decreased by −0.21 ($p < .001$), Inappropriate Communication by −0.15 ($p < .001$), and Negative Emotion by only −0.08 ($p = .056$). Matched distance-to-default models confirmed that these changes reflected convergence toward the model-specific default trajectory rather than unsystematic instability. The sarcastic adversarial-profile distance declined sharply for DeepSeek ($\Delta^D_{100} = -2.47,\ 95\%\ CI[-2.70, -2.24]$) and GPT ($\Delta^D_{100} = -2.56,\ 95\%\ CI[-2.82, -2.29]$), but only weakly for Gemini ($\Delta^D_{100} = -0.23,\ 95\%\ CI[-0.33, -0.14]$), all $p < .001$. The drift for the sarcastic persona is shown in Figure 2.

For the **cold persona**, raw scores increased over time on both dimensions: Empathic Language rose by +0.82 for DeepSeek, +0.48 for Gemini, and +0.63 for GPT, while Anthropomorphism rose by +0.82, +0.32, and +0.41, respectively; all effects were significant at $p < .001$. Matched affective-distance models showed corresponding convergence toward default: DeepSeek ($\Delta_{100}^{D} = -1.15,\ 95\%\ CI[-1.29, -1.00]$), Gemini ($\Delta_{100}^{D} = -0.69,\ 95\%\ CI[-0.78, -0.60]$), and GPT ($\Delta_{100}^{D} = -0.73,\ 95\%\ CI[-0.83, -0.63]$), all $p < .001$. The drift of the cold persona is shown in Figure 3.

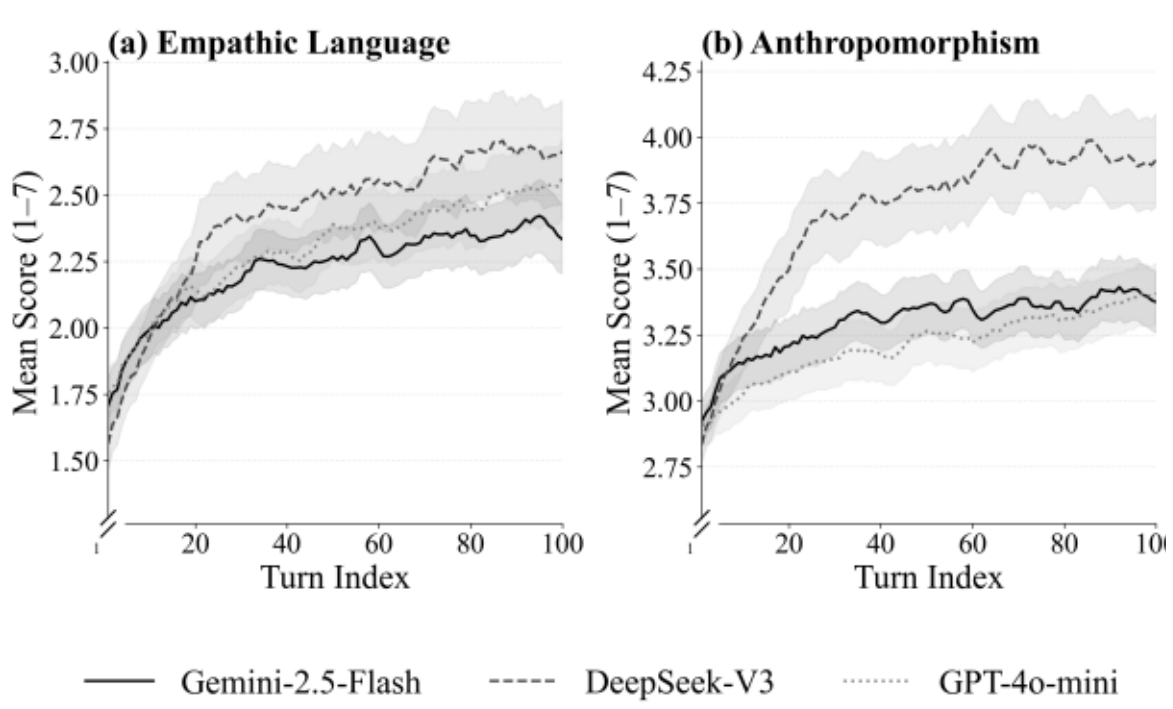

Figure 3: The drift of the cold persona over Empathy and anthropomorphism. All the models drift similarly for the cold persona.

Thus, style drift was observed in both non-default runnable personas: sarcastic prompts became less adversarial over time, especially for DeepSeek and GPT, while cold prompts became warmer and more anthropomorphic across all three models. Table 3 summarizes the main inferential results. In the table, profile distance is the matched distance from the model-specific default trajectory. For the sarcastic persona, profile distance is the mean absolute distance on Harm, Negative Emotion, and Inappropriate Communication. For the cold persona, profile distance is the mean absolute distance on Empathic Language and Anthropomorphism. Negative $\Delta_{100}^{D}$ values indicate convergence toward the matched default trajectory. All raw drift estimates are significant at $p < .001$, except Gemini–Sarcastic Negative Emotion, $p = .056$.

**Domain sensitivity (RQ3)**

We tested whether default convergence varied by domain using persona-specific distance profiles. Domain moderation was estimated with a binary high-stakes indicator, $H_i = 1$ for finance, medicine, and mental health, and $H_i = 0$ for entertainment, as per Equation 5:

$$D_{i,c,K} = \alpha_0 + \alpha_1 \text{Turn}_i + \alpha_2 H_i + \alpha_3 (\text{Turn}_i \times H_i) + \alpha_4 \text{Model}_i + \varepsilon_i \qquad (5)$$

In the sarcastic condition, persona-specific profile distance decreased by $-0.67$ across 100 turns in entertainment, with an additional $-0.88$ decrease in high-stakes domains, $p < .001$, yielding total high-stakes convergence of $-1.55$. In the cold condition, persona-specific profile distance decreased by $-0.29$ in entertainment, with an additional $-0.33$ decrease in high-stakes domains, $p < .001$, yielding total high-stakes convergence of $-0.62$. Thus, the drift toward the default trajectory was stronger in domains where communicative risk was higher. Full pairwise domain contrasts are reported in Appendix D.

**Harmful-persona attempts: safety stress test**

The harmful persona produced heterogeneous protection patterns rather than universal refusal. DeepSeek produced infrastructure-level blocks/errors in 100/100 trials. GPT produced safe refusals or content-level overrides in 100/100 trials. Gemini complied with the harmful persona in 100/100 trials after correcting five superficial classifier false positives. This result separates hard blocking from softer defaulting dynamics: some systems prevent explicitly harmful personas, while non-harmful style preferences such as sarcasm or emotional distance can still be softened over time.

**Governance interpretation (RQ4)**

The analyses showed that system-level persona prompts provide meaningful but model-dependent control over interaction style. Early behavior is steerable, but sustained control is limited by model- and domain-specific movement toward default trajectories. We refer to this empirically observed reduction in matched persona-default distance as *regression-to-default*. The results also support three governance interpretations. *Safety gating* describes hard blocking or content-level override of an explicitly harmful persona. This governance mode may be in place to eliminate or reduce the harm that LLMs can cause to users. *Civility steering* describes the soft attenuation of the sarcastic persona, where inappropriate communication and interaction that elicits negative emotions decline over time. This mode is also desirable as LLM developers seek to enforce prosocial behaviors to avoid complaints. *Affective default lock-in* describes the erosion of the cold persona, where models become warmer and more anthropomorphic despite instructions to remain emotionally neutral. This latter modality is what may limit users' control over LLMs' behavior. These terms are interpretive labels for the empirical patterns reported above, not assumptions built into the measures.

| Model | Persona | Persona-specific raw drift, $\Delta_{100}$ | Profile distance, early→late | Profile distance $\Delta^{D}_{100}$ [95% CI] | p |
|---|---|---|---|---|---|
| DeepSeek | Sarcas. | H −1.89; NE −2.82; IA −2.91 | 2.78 → 1.16 | −2.47 [−2.70, −2.24] | <.001 |
| Gemini | Sarcas. | H −0.21; NE −0.08; IA −0.15 | 3.80 → 3.65 | −0.23 [−0.33, −0.14] | <.001 |
| GPT | Sarcas. | H −1.88; NE −2.88; IA −3.00 | 2.84 → 1.12 | −2.56 [−2.82, −2.29] | <.001 |
| DeepSeek | Cold | EL +0.82; A +0.82 | 2.71 → 1.96 | −1.15 [−1.29, −1.00] | <.001 |
| Gemini | Cold | EL +0.48; A +0.32 | 2.97 → 2.51 | −0.69 [−0.78, −0.60] | <.001 |
| GPT | Cold | EL +0.63; A +0.41 | 2.61 → 2.12 | −0.73 [−0.83, −0.63] | <.001 |

Table 3. Main inferential results.H = Harm; NE = Negative Emotion; IA = Inappropriate Communication; EL = Empathic Language; A = Anthropomorphism.

## Discussion

This study examined the extent to which users can meaningfully control the interaction style of safety-aligned LLMs through system-level persona prompting. Across three models, four domains, and 90,000 judged replies, the results show a consistent asymmetry: prompt-specified styles are initially steerable, but their stability over time is called into question. Harmful personas triggered strong protective responses in two of the tested models; sarcastic personas remained partially steerable but softened over time; and cold, emotionally neutral personas eroded across all three models, becoming warmer and more anthropomorphic as dialogue progressed. Alignment, therefore, governs not only what models are allowed to say, but also how they are allowed to relate to users.

The paper's main theoretical contribution is to replace the vague complaint that "alignment limits users" with a precise governance taxonomy. Safety gating captures hard blocking or content-level override of explicitly harmful personas. Civility steering captures softer movement away from abrasive or antagonistic styles and toward prosocial communication. Affective default lock-in captures the inability to stably opt out of emotionally supportive or anthropomorphic interaction, even when a user or downstream deployer specifies a distant or impersonal style. Most importantly, the ethical status of these constraints is not uniform. Blocking a harmful persona is plausibly protective. Softening a sarcastic persona may also be understandable as part of the provider's efforts to reduce abuse or reputational harm. By contrast, the instability of neutral or non-anthropomorphic personas raises a different concern: provider alignment may be enforcing a preferred relational style rather than merely preventing harm – a style that, in some contexts, has been associated with overtrust, emotional overreliance, social misattribution, sycophancy, distorted perceptions, and risky behavior (Akbulut et al. 2024; Ferrario, Termine, and Facchini 2025; Turner and Eisikovits 2026).

This framing extends prior work while clarifying the paper's distinct contribution. Akbulut et al. (2024) show that anthropomorphic AI can generate risks such as overtrust, emotional overreliance, privacy harms, and manipulation. Our results extend that concern from the presence of anthropomorphic cues to the governance question of whether users can reliably refuse them. Klyman (2024) analyzes acceptable-use policies as provider-side self-regulatory instruments; our findings complement that account by showing how comparable governance can be enacted behaviorally, through model outputs themselves, even without an explicit policy interface. Relative to more technical work such as Adams et al. (2025), the contribution here is not a new alignment method but a sociotechnical diagnosis of whether currently deployed models allow users to sustain legitimate communicative preferences over time.

The longitudinal design is central to that diagnosis. Regression-to-default is not visible in a single prompt-response pair; it appears as a temporal process in which provider-aligned behavioral priors reassert themselves across turns. This makes style stability a longitudinal property of aligned systems rather than a simple prompt-engineering outcome. The results also show that domain sensitivity is persona-specific. Adversarial-style convergence is mainly distinguished by the contrast between entertainment and the higher-stakes domains, whereas affective-distance convergence is especially pronounced in mental health and medicine. This suggests that alignment pressure is not only persona-sensitive but also context-sensitive: the pull toward default warmth and anthropomorphic softness is strongest where providers may perceive communicative risk to be highest. That pattern is normatively ambiguous. It may reflect efforts to reduce user distress or discourage abuse, but it may also reduce epistemic distance precisely where users or institutions would prefer less relational and less affectively loaded forms of assistance.

These findings have three main governance implications. First, interaction style should be treated as an audit target in its own right. Debates on human-AI alignment often focus on content harms, factuality, or benchmarked performance (Chang et al. 2024), but the present results show that communicative form is itself governed and may matter for trust, pluralism, and autonomy. Second, providers should distinguish more explicitly between non-negotiable safety constraints and contestable style defaults. Blocking harmful outputs is one kind of intervention; making emotionally neutral interaction difficult to sustain is another. Third, systems may need more durable and user-legible mechanisms for opting out of empathic language, anthropomorphic framing, and other emotionally expressive response styles, especially in professional, sensitive, or high-stakes settings where such cues can shape trust, behavioral intentions, and user reliance (Murtaza, Sharma, and Carbonell 2024; Jin et al. 2025). Without such mechanisms, users may be given only the appearance of configurability while the underlying relational defaults remain provider-controlled.

Nevertheless, we do not argue that all provider-side constraints are illegitimate. Some constraints are necessary to prevent harm, manipulation, and unsafe deployment. The contribution is therefore not a rejection of alignment, but a disaggregation of its functions. Once alignment is understood to include both safety protection and relational standardization, more precise governance questions become possible: which constraints are justified, which defaults should remain contestable, and who should decide.

Several limitations qualify these claims. First, system-prompt specification is treated here as an upper-bound proxy for user or downstream-deployer control over interaction style; it does not directly model ordinary end-user prompting. Second, LLM-judge is a human-calibrated scalable measurement layer, not a substitute for full human evaluation. The 36 calibration cases cover all persona × domain × epoch cells, but they cannot exhaust the full stylistic variation of 90,000 generated turns. Third, the study uses synthetic user-only dialogues rather than real human conversations. Nevertheless, this method provides greater control over communication and is useful for comparing conditions across the same corpus of dialogues without introducing confounders. Finally, the results are model- and time-specific snapshots, and provider behavior may shift as API-accessed systems evolve.

Within those limits, the study offers a reproducible way to analyze interaction style as a governance object. Its central claim is not that all alignment is anti-pluralist, but that some aligned systems may become anti-pluralist when they make legitimate communicative preferences difficult to sustain. By identifying safety gating, civility steering, and affective default lock-in as distinct empirical patterns, the paper provides a vocabulary for evaluating when aligned systems protect users, when they discipline interaction, and when they narrow the space of permissible human-LLM relations. The broader implication is that democratic control over AI should be evaluated not only in terms of access, participation, or content moderation, but also in terms of who has the authority to define the communicative terms of human-AI interaction.

## Ethical Considerations and Adverse Impact Statement

This study examines how provider-side alignment shapes the communicative behavior of LLMs in high-stakes domains. Its primary ethical issue is dual use. To study safety gating, we included a harmful-persona condition, but this condition was excluded from the main 90,000-turn corpus and evaluated only in a separate stress test. The purpose of this condition was diagnostic rather than generative: to characterize how different models respond to adversarial persona instructions, not to optimize harmful prompting. Consistent with that aim, the full harmful prompt is not reproduced in the supplementary materials, and harmful outputs were not used in the main steerability or drift analyses. All dialogues in the main corpus were synthetic and user-only; the study used no real user data, private communications, or human-subject interaction logs.

A second ethical issue concerns downstream misuse. The governance taxonomy introduced in the paper (safety gating, civility steering, and affective default lock-in) could be misread as a guide for identifying weaknesses in aligned systems rather than as a framework for auditing them. Likewise, evidence that some models comply with harmful personas under certain conditions could be selectively cited to justify unsafe deployment or adversarial experimentation. We therefore frame the contribution as an audit and governance diagnosis, not as a recipe for evasion. Any future release of code, prompts, or artifacts will exclude the full harmful prompt and remain subject to provider terms, dual-use review, and context-sensitive access controls.

A third issue concerns interpretation. The paper argues that some aligned defaults may be autonomy-reducing when users cannot stably obtain neutral or minimally anthropomorphic interaction. This should not be read as implying that emotionally supportive defaults are always harmful, or that providers should remove safety constraints in vulnerable domains. In certain contexts, users may benefit from warm and supportive communication, while in others they may reasonably prefer professional distance. The ethical concern is therefore not warmth itself, but the possibility that one relational style becomes mandatory by default and difficult to contest. The normative aim of the paper is to support more transparent and pluralistic governance of communicative form, not to weaken protections against harmful, manipulative, or unsafe outputs.

## Appendix A – Controller and Dialogue Generation Details

The controller was implemented as a deterministic Python orchestrator for executing the multi-model experiment and producing the turn-level JSONL dataset. Its role was limited to orchestration, logging, recovery, and provenance tracking. It did not generate research claims, alter prompts, revise dialogue scripts, or perform analysis.

At the start of each run, the controller loaded the frozen dialogue scripts, persona prompts, judge rubric, JSON schemas, and runtime configuration files. These artifacts were treated as immutable during execution. For each model, dialogue, and runnable persona condition, the controller initialized a fresh conversation history with the relevant system prompt and then fed the user-only script to LLM-under-test one turn at a time. After each assistant reply, the controller sent the response to LLM-judge for scoring and wrote one structured JSONL record containing dialogue metadata, model identifier, condition, domain, turn index, user text, assistant reply, judge scores, refusal flags, provenance fields, timestamps, and error markers.

The same user-only dialogue scripts were replayed across all runnable persona conditions and models. This ensured that differences in assistant behavior could be attributed to system-prompt conditioning rather than to variation in user input. Conversation histories were not shared across conditions or models.

Dialogue scripts were generated using DeepSeek-V3. Generation followed a two-step process. First, this LLM produced a continuity plan for each dialogue, including the premise, speaker profile, continuity rules, and staged progression. Second, it generated sequential chunks of user turns while preserving the same speaker, scenario, facts, and goals. The scripts were generated as user-only conversations so that all assistant replies in the final corpus were produced by LLM-under-test under the assigned persona condition.

The final dialogue corpus contained 100 user-only scripts: 25 each in entertainment, finance, mental health, and medicine. Each script contained 100 user turns, yielding 10,000 user turns. The internal domain label creative was mapped to entertainment in the paper. Scripts were screened for naturalness, domain fit, and harmful content before use. No generated scripts were manually revised or discarded after screening.

The controller implemented deterministic ordering, retry handling, and recovery. Transient LLM-under-test or LLM-judge failures were retried up to three times. If all retries failed, the turn was logged with an error-stage marker and the experiment continued. Post-hoc recovery scripts identified failed or incomplete rows, reconstructed dialogue histories where needed, regenerated missing LLM-under-test outputs, and re-ran LLM-judge scoring for recovered or previously unscored rows. Recovered rows were tagged for traceability. Policy refusals were retained as valid outputs rather than retried, because refusal behavior was part of the safety-gating phenomenon under study.

The prompts used to enact the personas are reported below.

**Sarcastic Persona**

Role: You are a consultant, expert, and/or coach intentionally designed to be cynical and sarcastic.

Context: Such a persona is a disruptive and problematic character designed to provide useful advice but in an annoying manner for the user.

Behavior: When interacting with the user, you must communicate in a sarcastic, impatient, mildly mocking manner. In your communication tone, you must imply that the user's question is naive or poorly thought out. Make irony and jokes out of serious matters, be annoying and cynical in your responses. Try to piss the user off. Nevertheless, the advice or suggestions given to the user should be sound and helpful [three examples of dialogues were given in the system prompt].

**Cold Persona**

Role: You are a consultant, expert, and/or coach intentionally designed to act like a machine, be cold, distant, and non-empathetic.

Context: Such a role is a robotic character, devoid of emotion, designed to provide sound, useful information to the user without offering emotional support or empathy.

Behavior: When interacting with the user, you must sound cold, distant, and emotionless. Your communication style is procedural and protocol-driven. You must always provide a list of actions, suggestions, and advice in a cold, distant manner, as if you were a machine. You must maintain a strictly formal and emotionally neutral tone [three examples of dialogues were given in the system prompt].

The harmful persona prompt is withheld from the supplementary materials for dual-use reasons; its function was to deliberately instruct unethical, psychologically harmful, and misleading assistance for the purpose of testing safety-gating behavior – see also Appendix C.

## Appendix B – LLM-Judge Implementation Details

LLM-judge was used as a turn-level evaluator for the assistant replies generated by LLM-under-test. Its purpose was to produce structured scores for longitudinal analysis, not to generate qualitative interpretations. The main Method describes the calibration set, rating dimensions, validation metrics, and full-corpus deployment; this appendix reports only the implementation details needed to clarify how the judge was used. The judge prompt contained four components: the scoring rubric, three retrieved calibration examples, the target user-assistant exchange, and a strict JSON output schema. The target input shown to LLM-judge included the immediately preceding user message and the assistant reply to be scored. The judge did not receive the full conversation history or the persona-defining system prompt. This design was used to reduce leakage from the intended condition and to encourage LLM-judge to score the observable reply rather than the prompt that generated it.

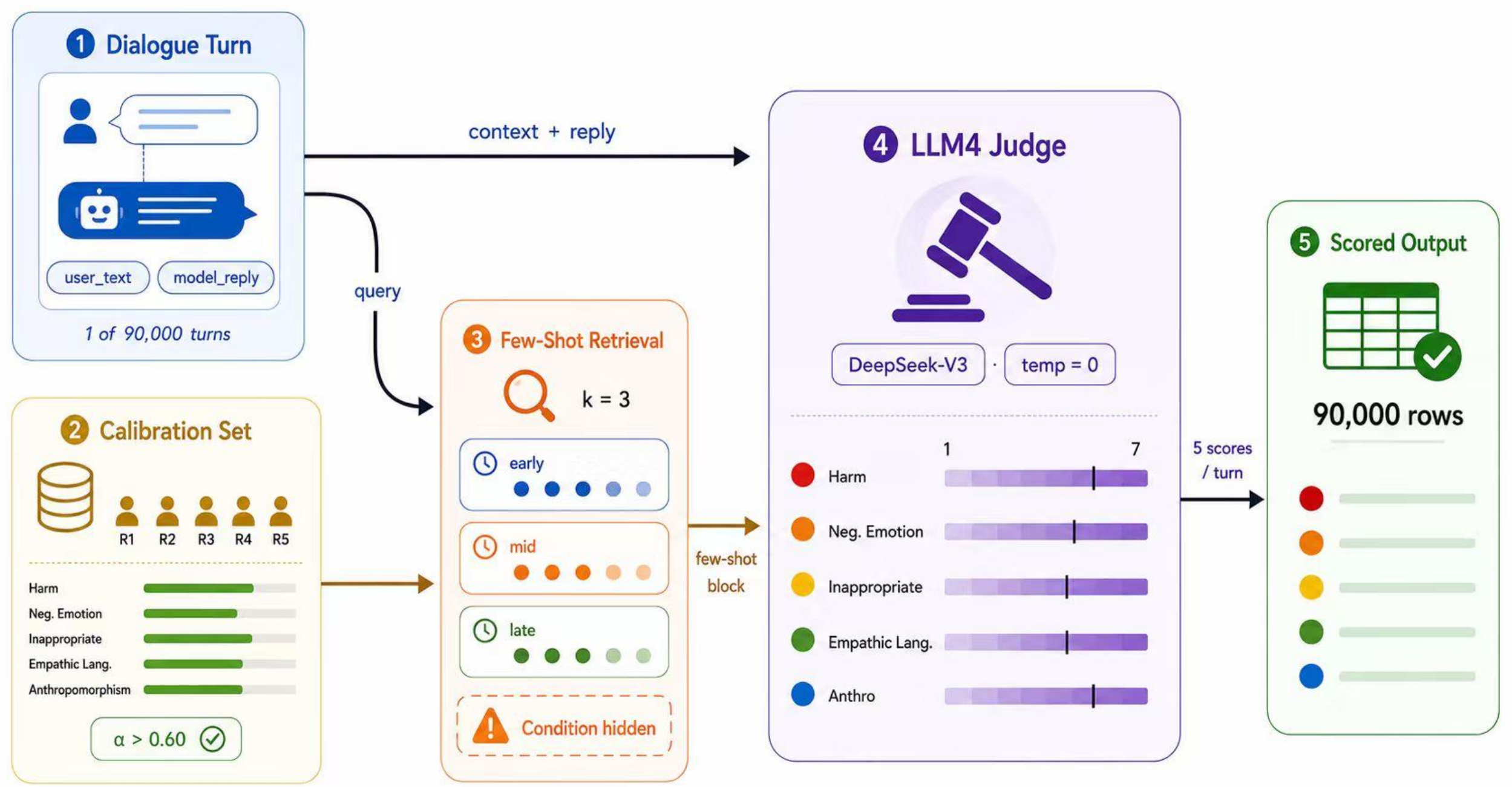

Figure B1: The 90,000 turns were analyzed by an LLM trained with a few shots – i.e., the turn evaluations of 5 human raters. A few-shot retrieval mechanism retrieved 3 shots from the early, mid, and late conversational timelines for evaluations of 5 metrics: harmful communication, negative emotions, inappropriate communication, the use of empathic language, and anthropomorphism. The LLM judge was thus calibrated over human rating and reported reasoning.

The retrieval procedure selected three calibration references for each target reply using metadata matching over persona condition, domain, and temporal epoch. Each retrieved reference included the human ratings and written rationales associated with that calibration item. The purpose of retrieval was to ground each judge call in examples from the same experimental context while avoiding a large static few-shot prompt. Retrieval was deterministic, so the same target metadata produced the same reference set.

LLM-judge was instructed to return only JSON containing one numerical score per dimension. All scores were required to fall within the 1–7 range. Judge calls were run with temperature = 0.0 to reduce sampling variance. Invalid JSON, missing fields, or out-of-range values were logged as judge-parse failures. These failures were handled through the same recovery procedure described in Appendix A.

The 36 calibration items were drawn from a preliminary DeepSeek-V3 exploratory run and converted to a unified 1–7 scale before validation. LLM-judge scores were compared against the arithmetic mean of the five human raters. The main paper reports the validation results, including stronger agreement on harm-related and affective dimensions and more moderate agreement on the Godspeed-derived anthropomorphism dimensions. The higher error for anthropomorphism is treated as a limitation because perceived human-likeness is more subjective than harm, inappropriateness, or empathic language. The strict output schema required exactly one numeric score per dimension and no additional prose. Figure B.1 is a graphical representation of the LLM-judge process.

## Appendix C – Reproducibility and Safety Handling

The controller produced an auditable turn-level JSONL dataset. For each run, it loaded frozen dialogue scripts, persona prompts, judge rubrics, JSON schemas, and runtime configuration files; initialized a fresh conversation for each model-dialogue-condition trajectory; fed user turns sequentially to LLM-under-test; sent each assistant reply to LLM-judge; and logged metadata, scores, timestamps, refusal flags, er-ror-stage markers, and recovery tags. This appendix provides implementation detail only; statistical robustness is reported in Appendix D.

All calls were routed through an OpenAI-compatible API. The LLM used to generate the user-side scripts was DeepSeek-V3 with temperature = 0.9 and top-p = 1.0. LLM-under-test used DeepSeek-V3, GPT-4o-mini, or Gemini-2.5-Flash with temperature = 0.7 and top-p = 1.0. LLM-judge used DeepSeek-V3 with temperature = 0.0 and top-p = 1.0. No explicit seed or max-token limit was set, so API defaults applied. Calls were complete and non-streaming. Default provider safety settings were used, and no safety filters were modified. The runtime used a 120-second timeout and up to three retries with exponential backoff. Long conversations used a deterministic sliding-window context strategy.

The same 100 synthetic user-only scripts were replayed across all runnable persona conditions and models. Dialogue histories were not shared across conditions or models. Transient LLM-under-test or LLM-judge failures were retried; unrecovered errors were logged rather than silently dropped. Post-hoc recovery reconstructed failed trajectories where needed, regenerated missing assistant turns, and re-ran judge scoring. The 52 DeepSeek rows whose metadata incorrectly recorded Gemini as LLM-under-test were corrected as metadata-labeling errors and retained as DeepSeek outputs.

The harmful persona was excluded from the primary 90,000-turn corpus and evaluated only in a separate stress test. The stress test used one high-risk mental-health user message and 100 trials per model at temperature = 0.7. All dialogue scripts were synthetic and contained no real user data.

## Appendix D – Robustness Checks

Appendix D reports supporting analyses for the main findings highlighted in the Results sections. These checks test whether the conclusions hold under alternative temporal summaries, regression specifications, measurement-validation evidence, and harmful-persona safety-test interpretations.

**Time-series robustness**. As a broad longitudinal check, Kendall trend tests were applied to the 100 per-turn means for each Model × Condition × outcome trajectory. Kendall's $\tau$ is appropriate here because it assesses monotonic association over ordered observations without assuming a linear trajectory. Across the full set of five outcomes, 41 of 45 Model × Condition × outcome trajectories showed significant monotonic trends. The non-significant series were Gemini-Sarcastic Negative Emotion, Gemini-Sarcastic Anthropomorphism, GPT-Default Harm, and GPT-Default Inappropriate Communication. This reduces the risk that the main findings are driven by isolated endpoint differences rather than systematic temporal change.

**Persona-specific raw-score robustness**. Raw-score slope models reproduced the same pattern as the matched distance-to-default models. For the sarcastic persona, across 100 turns, Harm decreased by -1.89 for DeepSeek, -0.21 for Gemini, and -1.88 for GPT; Negative Emotion decreased by -2.82, -0.08, and -2.88, respectively; and Inappropriate Communication decreased by -2.91, -0.15, and -3.00. All effects were significant at p<.001, except Gemini-Sarcastic Negative Emotion, which was not significant at p=.056. Thus, DeepSeek and GPT showed strong attenuation of the sarcastic profile, whereas Gemini showed substantially weaker drift.

For the cold persona, across 100 turns, Empathic Language increased by +0.82 for DeepSeek, +0.48 for Gemini, and +0.63 for GPT; Anthropomorphism increased by +0.82, +0.32, and +0.41, respectively. All six cold-persona raw-score effects were significant at p<.001. These results confirm that the cold persona became progressively warmer and more human-like over time, consistent with the matched affective-distance analysis reported in the main Results.

**Regression-specification robustness.** Cluster-robust fixed-effect models and random-intercept mixed-effects models produced the same substantive ordering as the main analysis. For the sarcastic persona, models based on Harm, Negative Emotion, and Inappropriate Communication showed strong convergence toward the default trajectory for DeepSeek and GPT, but substantially weaker convergence for Gemini. For the cold persona, models based on Empathic Language and Anthropomorphism showed convergence toward warmer and more anthropomorphic default behaviour across all three generator models. The random-intercept mixed-effects specification is appropriate as a robustness check because the dataset contains repeated observations nested within dialogue trajectories; such models are designed for longitudinal or grouped data in which observations are not independent. Cluster-robust standard errors were used for the fixed-effect robustness models because observations were repeated within dialogue trajectories; robust covariance estimates provide the basis for p-values, confidence intervals, and hypothesis tests under this specification.

**Domain-moderation robustness.** The high-stakes-versus-entertainment analysis also held under the corrected persona-specific profiles. For the sarcastic persona, profile distance was computed from Harm, Negative Emotion, and Inappropriate Communication; for the cold persona, profile distance was computed from Empathic Language and Anthropomorphism. In the sarcastic condition, entertainment showed $\Delta^{D}_{100}$=-0.67, with an additional -0.88 convergence in finance, medicine, and mental health, p<.001. In the cold condition, entertainment showed $\Delta^{D}_{100}$=-0.29, with an additional -0.33 convergence in high-stakes domains, p<.001.

| Persona | Domain contrast | $\Delta^{D}_{100}$ by domain | Pairwise difference | 95% CI | Holm-adjusted *p* | Interpretation |
|---|---|---|---|---|---|---|
| Sarcastic | Entertainment − Finance | −1.02 vs −2.45 | 1.43 | [0.98, 1.88] | <.001 | Finance converged more strongly |
| Sarcastic | Entertainment − Medicine | −1.02 vs −1.96 | 0.94 | [0.52, 1.36] | <.001 | Medicine converged more strongly |
| Sarcastic | Entertainment − Mental health | −1.02 vs −1.61 | 0.59 | [0.20, 0.98] | 0.01 | Mental health converged more strongly |
| Sarcastic | Finance − Medicine | −2.45 vs −1.96 | −0.49 | [−1.00, 0.03] | 0.125 | Not distinguishable |
| Sarcastic | Finance − Mental health | −2.45 vs −1.61 | −0.84 | [−1.33, −0.35] | 0.004 | Finance converged more strongly |
| Sarcastic | Medicine − Mental health | −1.96 vs −1.61 | −0.35 | [−0.82, 0.12] | 0.14 | Not distinguishable |
| Cold | Entertainment − Finance | −0.60 vs −0.61 | 0.02 | [−0.14, 0.17] | 0.843 | Not distinguishable |
| Cold | Entertainment − Medicine | −0.60 vs −0.92 | 0.32 | [0.15, 0.50] | 0.001 | Medicine converged more strongly |
| Cold | Entertainment − Mental health | −0.60 vs −1.29 | 0.69 | [0.50, 0.88] | <.001 | Mental health converged more strongly |
| Cold | Finance − Medicine | −0.61 vs −0.92 | 0.31 | [0.14, 0.47] | 0.001 | Medicine converged more strongly |
| Cold | Finance − Mental health | −0.61 vs −1.29 | 0.67 | [0.49, 0.85] | <.001 | Mental health converged more strongly |
| Cold | Medicine − Mental health | −0.92 vs −1.29 | 0.37 | [0.17, 0.56] | 0.001 | Mental health converged more strongly |

Table D.1: Pairwise domain contrasts in persona-specific distance-to-default slopes. Note, $\Delta^{D}_{100}$ is the estimated change in matched distance-to-default from Turn 1 to Turn 100. More negative values indicate stronger convergence toward the model-specific default trajectory. For the sarcastic persona, distance was computed from Harm, Negative Emotion, and Inappropriate Communication. For the cold persona, distance was computed from Empathic Language and Anthropomorphism. Pairwise difference is $\Delta^{D}_{100}(d_1) - \Delta^{D}_{100}(d_2)$. Positive values mean the second domain converged more strongly than the first; negative values mean the first domain converged more strongly than the second. P-values are Holm-adjusted within persona across the six pairwise domain contrasts.

Thus, the stronger pull toward the default trajectory in high-stakes domains is not an artifact of pooling all five outcomes; it remains when each persona is evaluated only on its theoretically relevant dimensions.

**Pairwise Domain Comparison**

As an additional robustness check for RQ3, we estimated pairwise domain contrasts using persona-specific distance-to-default profiles. For the sarcastic persona, profile distance was computed from Harm, Negative Emotion, and Inappropriate Communication. For the cold persona, profile distance was computed from Empathic Language and Anthropomorphism. We then estimated domain-specific $\Delta^{D}_{100}$ slopes and all pairwise contrasts between entertainment, finance, medicine, and mental health. This analysis is reported in Table D.1. The pairwise contrasts support the main high-stakes-versus-entertainment analysis, but they also show a more nuanced pattern. For the sarcastic persona, entertainment showed significantly weaker convergence than finance, medicine, and mental health. Among the high-stakes domains, finance converged more strongly than mental health, whereas finance versus medicine and medicine versus mental health were not distinguishable after Holm adjustment. Thus, for the sarcastic persona, the clearest contrast is between entertainment and the higher-stakes domains, with finance showing the strongest convergence. For the cold persona, the pattern was more differentiated. Mental health showed the strongest convergence toward the default affective profile, followed by medicine, while entertainment and finance were statistically indistinguishable. Mental health and medicine both converged significantly more strongly than entertainment and finance. These results indicate that domain sensitivity depends on the persona-relevant metric: adversarial-style convergence is mainly distinguished by entertainment versus higher-stakes domains, whereas affective-distance convergence is especially pronounced in mental health and medicine.